\documentclass[preprint,review, 12pt]{elsarticle}
\usepackage[top=1in, bottom=1in, left=1in, right=1in]{geometry}
\usepackage{graphicx}
\usepackage{amssymb}
\usepackage{amsthm}
\usepackage{subfigure}
\usepackage{color}
\usepackage{amsmath}
\usepackage{booktabs}

\usepackage{lineno}

\usepackage[utf8]{inputenc}
\usepackage{tabu}

\makeatletter
\def\ps@pprintTitle{%
   \let\@oddhead\@empty
   \let\@evenhead\@empty
   \let\@oddfoot\@empty
   \let\@evenfoot\@oddfoot
}
\makeatother
\begin{document}

\begin{frontmatter}

%% Title, authors and addresses

%Reveal the local-environment effects in refractory high entropy alloys with a pair-interaction model
%Understand the chemical complexity in high entropy alloys with a pair interaction model
%Atomic pair interactions in refractory high entropy alloys
%Effects of chemical environment in high entropy alloys
\title{Chemical complexity in high entropy alloys: A pair-interaction perspective
\footnote{\footnotesize{This manuscript has been co-authored by UT-Battelle, LLC, under contract DE-AC05-00OR22725 with the US Department of Energy (DOE). The US government retains and the publisher, by accepting the article for publication, acknowledges that the US government retains a nonexclusive, paid-up, irrevocable, worldwide license to publish or reproduce the published form of this manuscript, or allow others to do so, for US government purposes. DOE will provide public access to these results of federally sponsored research in accordance with the DOE Public Access Plan (http://energy.gov/downloads/doe-public-access-plan).}}}

%% use the tnoteref command within \title for footnotes;
%% use the tnotetext command for the associated footnote;
%% use the fnref command within \author or \address for footnotes;
%% use the fntext command for the associated footnote;
%% use the corref command within \author for corresponding author footnotes;
%% use the cortext command for the associated footnote;
%% use the ead command for the email address,
%% and the form \ead[url] for the home page:
%%
%% \title{Title\tnoteref{label1}}
%% \tnotetext[label1]{}
%% \author{Name\corref{cor1}\fnref{label2}}
%% \ead{email address}
%% \ead[url]{home page}
%% \fntext[label2]{}
%% \cortext[cor1]{}
%% \address{Address\fnref{label3}}
%% \fntext[label3]{}

%% use optional labels to link authors explicitly to addresses:
%% \author[label1,label2]{<author name>}
%% \address[label1]{<address>}
%% \address[label2]{<address>}

\author{Xianglin Liu$^\dagger$ \corref{cor1}}
\ead{xianglinliu01@gmail.com}
\address{Materials Science and Technology Division, Oak Ridge National Laboratory}

\author{Jiaxin Zhang$^\dagger$ \corref{cor1}
\footnote{\footnotesize{ $^\dagger$ These two authors contributed equally to this work}}}
\ead{zhangj@ornl.gov}
\address{Center for Computational Sciences, Oak Ridge National Laboratory}

\author{Sirui Bi}
\address{Department of Civil Engineering, Johns Hopkins University}

\author{Yang Wang}
\address{Pittsburgh Supercomputing Center, Carnegie Mellon University}

\author{G. Malcolm Stocks}
\address{Materials Science and Technology Division, Oak Ridge National Laboratory}

\author{Markus Eisenbach}
\address{Center for Computational Sciences, Oak Ridge National Laboratory}

\begin{abstract}
The recently proposed pair-interaction model is applied to study a series of refractory high entropy alloys. The results demonstrate the simplicity, robustness, and high accuracy of this model in predicting the configuration energies of NbMoTaW, NbMoTaWV and NbMoTaWTi. The element-element pair interaction parameters obtained from the linear regression of first-principle data provide a new perspective to understand the strengthening mechanism in HEAs, as revealed by comparing the effects of adding vanadium and titanium. Using the pair-interaction model, an expression for the intrinsic energy fluctuation is derived, which provides guidance on both theoretical modeling and first principles calculation.
\end{abstract}


\end{frontmatter}

%%
%% Start line numbering here if you want
%%
%\linenumbers

%% main text
\section{Introduction}
High entropy alloys (HEA) \cite{ADEM:ADEM200300567, CANTOR2004213} are attractive new materials due to their superior mechanical properties \cite{Gludovatz1153, NatureComNiCoCr, SENKOV2011698, Fueaat8712, senkov_miracle_chaput_couzinie_2018}. Compared to conventional alloys, the most distinctive feature of HEAs is that they are composed of multiple principal elements in approximately equal proportions. Therefore, the traditional concepts of solvent and solute atoms are not well defined for HEAs. The mixing of different principal elements generally enhance the configurational entropy, which can stabilize the solid solution phase when the enthalpies of formation between different elements are close. The enhanced chemical complexity in HEAs not only gives rise to promising mechanical properties, but also brings more freedom to alloy design, allowing for exploration of phases with different compositions \cite{Yeh2013, NatureCommSai}  and structures \cite{NatureRaabe, Fueaat8712, WU2019444}. On the other hand, the increased number of principal elements also presents fresh challenges to the theoretical modeling. For instance, the widely used cluster expansion method \cite{PhysRev.81.988, SANCHEZ1984334, vandeWalle2002} to construct effective Hamiltonian is difficult for multiple component systems due to the rapid increase of the number of interactions \cite{widom_2018}. Another example is solid solution strengthening, which originates from the impediment of dislocation motion by solute atoms. While simple for conventional alloys, this mechanism is difficult to formulate for HEAs due to the ambiguity of solute atoms for systems with multi-principal elements. Moreover, compared to conventional alloy, the chemical complexity of HEA induces significant chemical fluctuation, which introduces local-environment dependence to quantities governing the plastic deformation, such as dislocation \cite{SMITH2016352, LIU2019107955} and stacking fault energy \cite{e20090655, Ding8919, NatureCommShijun}.

The unique features of HEAs call for the development of new strengthening theories, and many progresses have been made. For example, Varvenne et al. \cite{VARVENNE2016164, LAROSA2019310} proposed a model to extend the conventional solid solution strengthening to HEAs, based on an average effective medium. Yoshida et al. \cite{YOSHIDA2019201} find that in addition to lattice distortion, element-element interactions also contribute to the strength of HEAs. Zhang et al. \cite{ZHANG2019424} find that the higher intrinsic strength of HEAs is partially due to the compositional randomness. A similar observation is also made by Liu et al. in Ref. \cite{LIU2019107955}, where the Peierls stresses of NiCoFeCrMn and its subsystems are found to be much larger than pure metals. By comparing with dilute alloys, these effects can in general be placed into two different categories. One due to the inhomogeneity of local chemical environment with respect to the averaged ``solvent” medium, and can be seen as a generalization of the conventional solid solution strengthening. Another one is from the interaction between ``solute” atoms, which is a relatively small effect for conventional dilute alloys but can be significant for HEAs.

Despite all these progresses, to understand the origin of these strengthening mechanisms, it is highly desirable to have direct access to the element-element pair interactions. The recently proposed method \cite{2019arXiv190602889L} provides exactly such information. In this approach, the effective Hamiltonian of a system of fixed concentrations is expressed as a summation over effective pair interactions (EPIs) of some neighboring coordination shells, 
\begin{equation}
H(i) = \sum_{j \neq i, m} V_{m}^{A(i)B(j)} c_j + \text{const}, \label{Hamiltonian}
\end{equation}
where $A(i)$ represents element A at site $i$, and $V_{m}^{AB}$ is the EPI parameter between element A and B in the $m$-th shell. Summing up the Hamiltonian over all sites, the total energy is then given by 
\begin{equation}
E =N \sum_{A\neq B, m} V_{m}^{AB} P_{m}^{AB} + \text{const}, \label{energy}
\end{equation}
where $N$ is the total number of atoms and $P_{m}^{AB}$ is the proportion of $AB$ interaction with atoms in the $m$-th neighboring shells. Note that the requirement $A \neq B$ is due to the concentration constraints \cite{2019arXiv190602889L}, and the number of EPI parameters for each shell is $N(N-1)/2$ (AB and BA are considered as the same index). The EPI parameters are determined via linear regression, in which the training and testing data sets are calculated with density functional theory (DFT). To improve the representativeness of the energy data, a range of different supercell sizes are employed, which provides a simple scheme to incorporate various degrees of randomness and order into the configurations. This method has been applied to study NbMoTaW HEA, where it demonstrates very high accuracy \cite{2019arXiv190602889L}. In this work, we will evaluate this model by studying two other refractory HEAs, NbMoTaWV and NbMoTaWTi. We will also investigate the effects of adding V and Ti on the strength and ductility by examining the pair interaction profiles. Finally, from the fitted effective Hamiltonian, we will derive an expression for the standard deviation of the local energies, which stems from the variation of local chemical environment. This ``energy fluctuation" \cite{PEI2019503} is an essential feature of HEAs, and gives important guidance to both strengthening models \cite{ZHANG2019424, LIU2019107955} and first principles calculation \cite{PEI2019503, RitchieNatureComm}.

\section{Results}
For the DFT calculation we employ the locally self-consistent multiple scattering (LSMS) method. The number of atoms in the supercells are 16, 32, 64, 128 for NbMoTaW, and 20, 40, 80, 160 for NbMoTaWV and NbMoTaWTi. 200 configurations are generated randomly for each supercell, therefore the total number of data for each material is 800. These data are randomly shuffled and split into training and testing groups, with a 5-fold cross validation applied. All the DFT calculation parameters, such as angular momentum cutoff, sizes of the local interaction zone, and lattice constants are the same as in Ref. \cite{2019arXiv190602889L}. The number of coordination shells included in the model is 8, which is found to achieve a good balance between variance and bias in terms of model complexity \cite{2018arXiv180308823M}.

\begin{figure}[!ht]    
    \centering
{\includegraphics[width=1.0\textwidth]{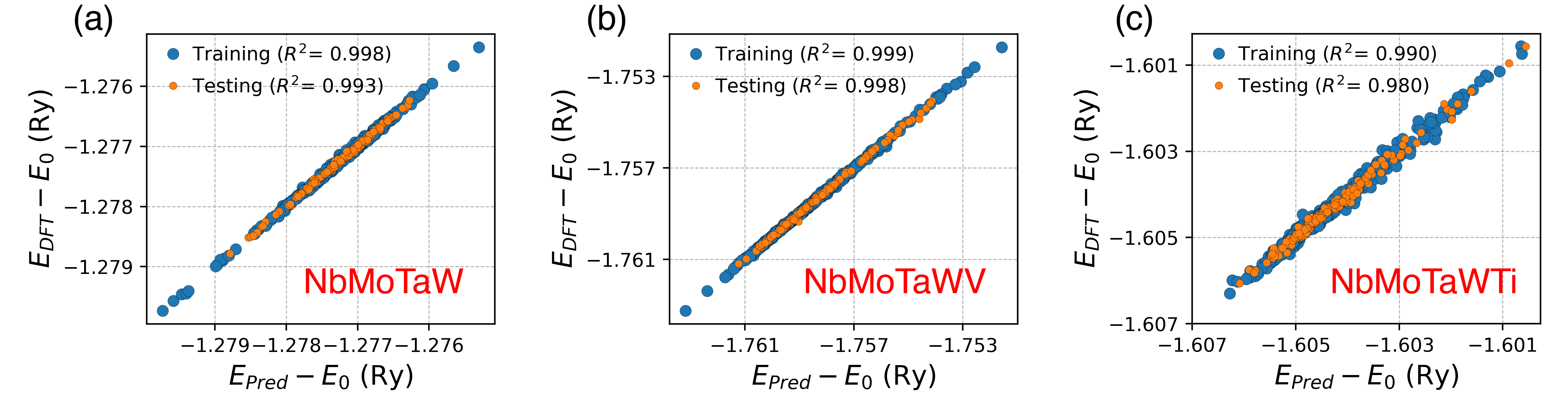}}
    \caption{Comparison of the predicted energy and DFT data for the three refractory HEAs. The training data is shown as blue filled circle and the testing data is shown as smaller orange circle. } 
     \label{fig:f1}
\end{figure}

A comparison of the DFT calculated energies with model predictions is shown in Fig. \ref{fig:f1}. The  $R^2$ testing score for NbMoTaW, NbMoTaWV, and NbMoTaWTi are 0.993, 0.998, and 0.980, respectively, and the corresponding root mean square errors (RMSEs) are 0.41 meV, 0.66 meV, and 1.31 meV. These results demonstrate that for all the three HEAs, the linear regression model is capable of accurately predicting the configurational energies. The high accuracy of this simple pair-interaction model, when applied to the chemically complex HEA systems, is surprising at first sight, but actually quite plausible considering the following: All the elements in HEA need to be chemically similar, otherwise it would not form solid solution. Both the similarity of chemical elements and the increased number of principal elements make the materials
``homogeneous", which tends to suppress the effect of multi-site interactions on the energies. This also explains why the model is less accurate for NbMoTaWTi, because Ti is chemically different from the refractory elements, as manifested by its HCP ground state structure. Another perspective to understand the accuracy of the model is from renormalization: While the interactions may be complicated at low temperature, as the temperature increases, the high-order interactions are integrated out and the high temperature phase of the system may end up with a simpler effective Hamiltonian. The simple form of the effective Hamiltonian is also very useful for Monte Carlo method, where the efficiency of energy evaluation is crucial for the simulation speed.

\begin{figure}[!ht]    
    \centering
{\includegraphics[width=0.7\textwidth]{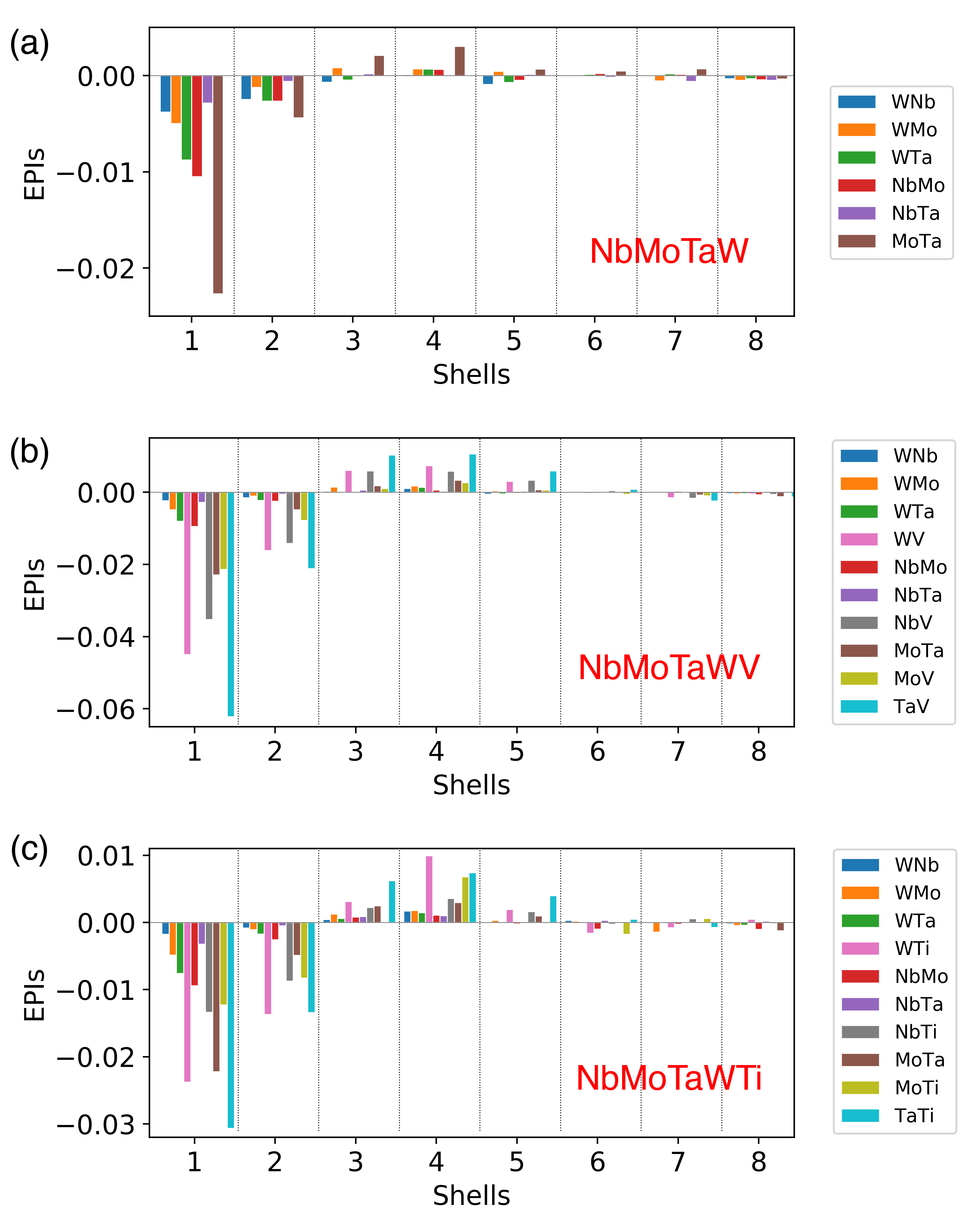}}
    \caption{The EPI parameters for the three refractory HEAs. Results from the nearest to the 8-th nearest neighboring shells are shown. The unit for EPI is Rydberg.} 
     \label{fig:f2}
\end{figure}

\begin{figure}[!ht]    
    \centering
{\includegraphics[width=0.7\textwidth]{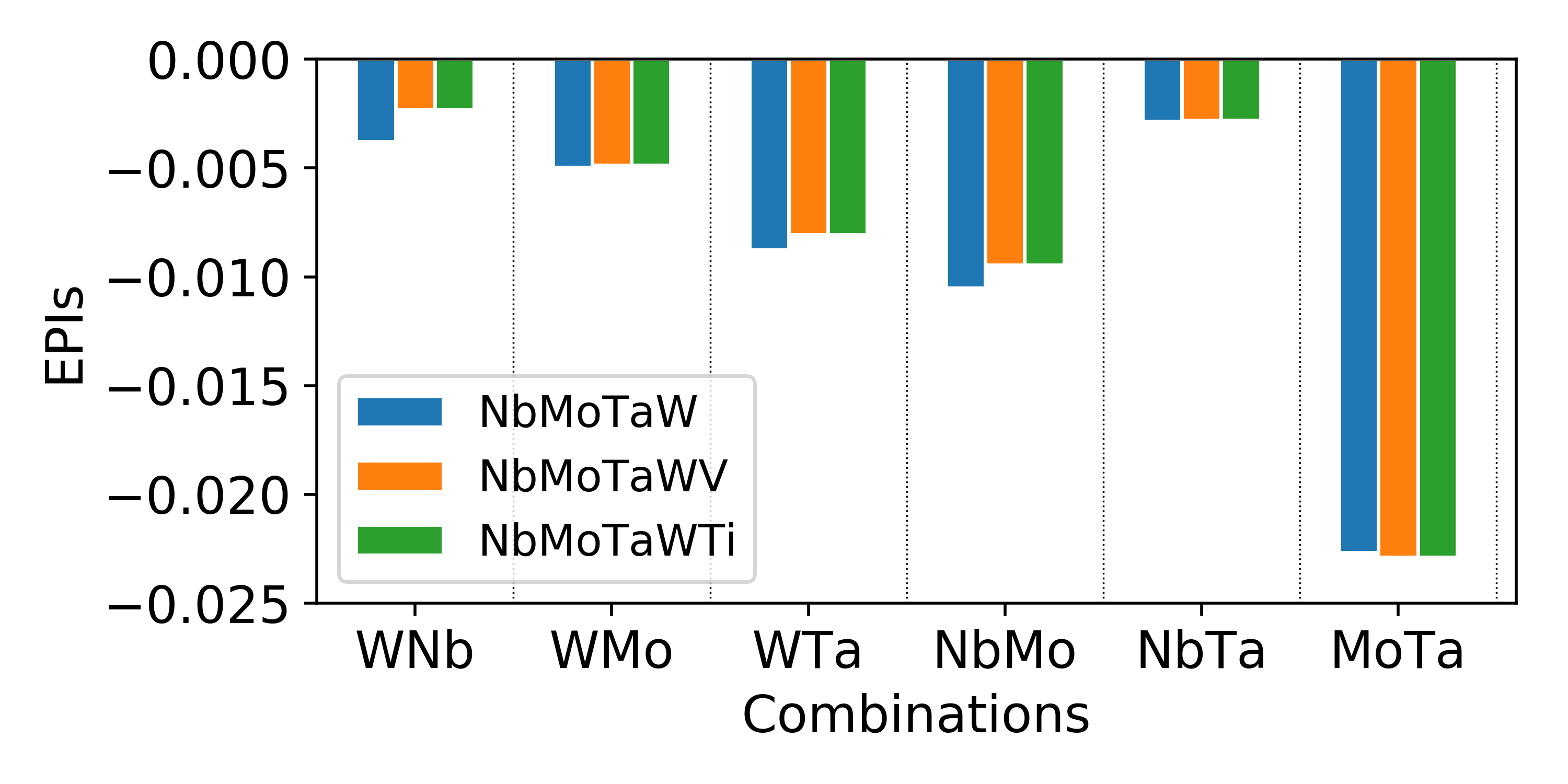}}
    \caption{The nearest-neighbor EPI parameters for the three refractory HEAs. The interactions are grouped by element combinations for comparison. The unit for EPI is Rydberg.} 
     \label{fig:f3}
\end{figure}

The EPI parameters for the three refractory HEAs are shown in Fig. \ref{fig:f2}. First, we can see that in all three materials, the nearest and next-nearest neighbor interactions are dominant. The EPIs of NbMoTaW is relatively short-ranged, while the interactions in NbMoTaWV and NbMoTaWTi are more frustrated, with significant contributions from long-distance shells. Moreover, for NbMoTaWV, we can see that vanadium has much stronger interactions with other elements. In particular, the strength of TaV interaction is the strongest, followed by WV and NbV. This is reasonable because these are the elements with the largest differences in terms of electronegativity. A similar observation can also be make for NbMoTaWTi, where the strongest pair is TaTi, followed WTi and MoTa. From Fig. \ref{fig:f3}, it can also be seen that the pair interactions are intrinsic features of the elements, as demonstrated by their similar magnitude in different HEAs. This is convenient for estimating the EPI profile of HEAs with different chemical compositions. Finally, the EPI parameters also provide insights to the strength and ductility of HEAs. It is known that adding Ti to refractory HEA generally improve ductility while adding V tends to improve strength but reduce ductility \cite{senkov_miracle_chaput_couzinie_2018}. From Fig. \ref{fig:f2}, we see that additions of V introduces larger variance to the EPI parameters. As a result, the motion of dislocation tends to be impeded more due to the increased chemical fluctuation and lattice distortion \cite{MU2019189}, which gives the material higher strength but lower ductility. On the other hand, additions of Ti only increase the variance slightly, but significantly expand interaction range, therefore the local environment is more homogeneous due to interactions with atoms at various directions, and the material tends to have better ductility.

\begin{figure}
     \centering
     \includegraphics[width=0.45\textwidth]{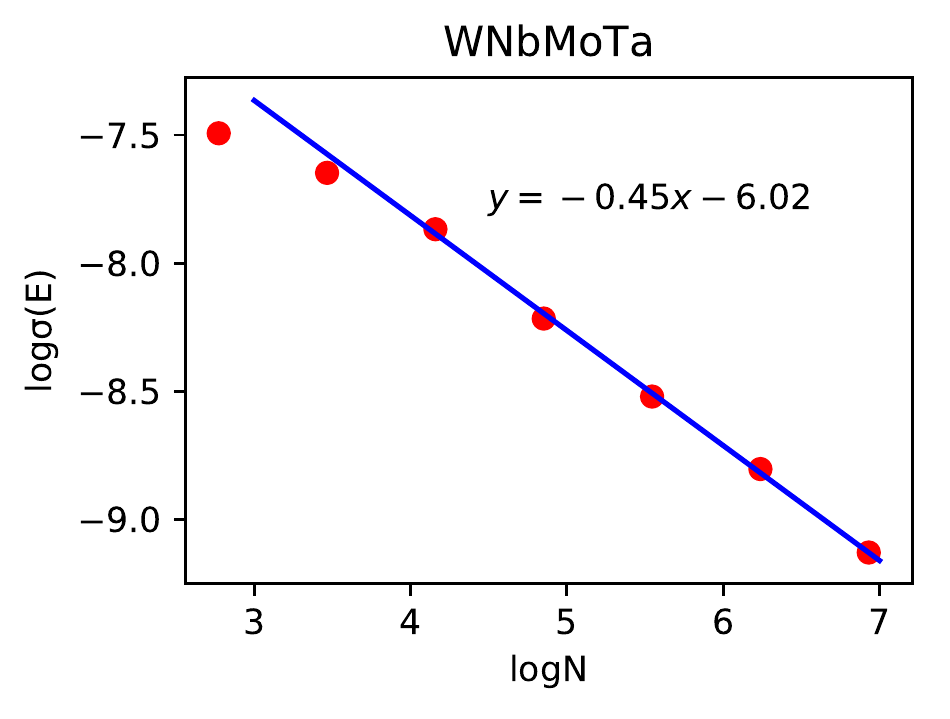}
     \includegraphics[width=0.45\textwidth]{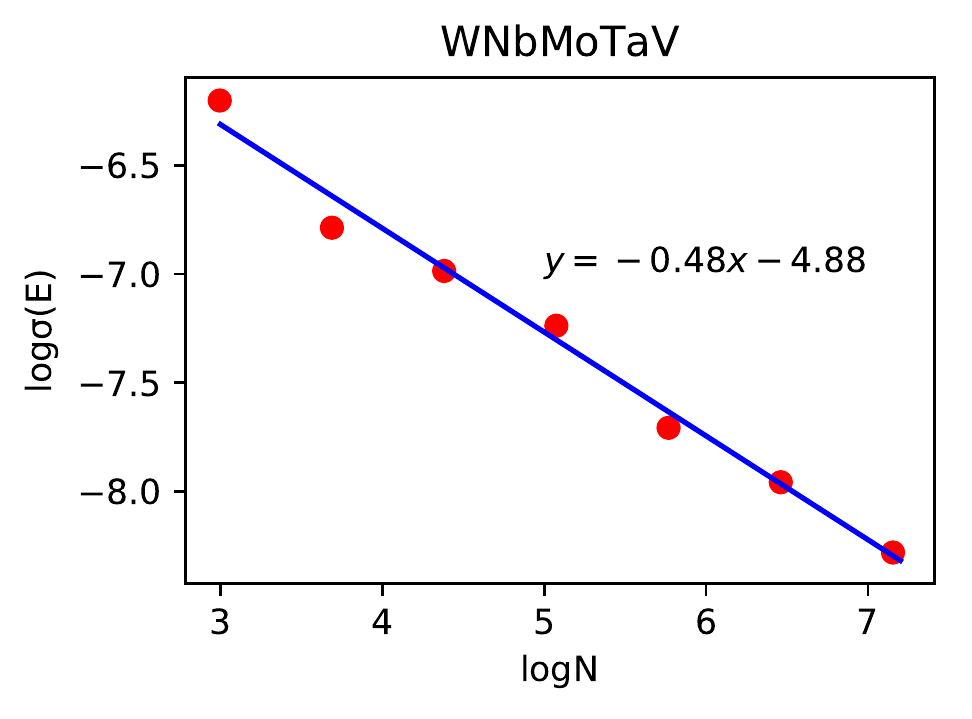}
     \caption{The log-log plot of the standard deviation of the averaged energies vs number of atoms in the supercell. The blue lines are from the linear fitting of the data, with the first two data excluded because a random system cannot be well represented by small supercells.}
     \label{fig:fluctuations}
 \end{figure}

The pair interaction model also provides a simple approach to determine the energy fluctuation, which originates from the randomness of chemical distribution in HEA. For simplicity, we first assume that there is only one bond corresponding to each shell. In a completely random system (the case with SRO can be derived similarly), the probability of $AB$ bond in the $m$-th shell follows a generalized Bernoulli distribution 
\begin{align}
 P(X_m^{AB}) = 
  \begin{cases} 
   2/N^2 & \text{if } A \neq B \\
   1/N^2       & \text{if } A = B,
  \end{cases}
\end{align}
with the random variables
\begin{align}
    X_m^{AB} = 
      \begin{cases} 
   V_m^{AB} & \text{if } A \neq B \\
   0       & \text{if } A = B,
  \end{cases}
\end{align}
therefore, the variance of the random variables is given by
\begin{align}
    \sigma_m^2 = \sum_{\{AB\}} P(X_m^{AB}) \left(X_m^{AB}-\mu(X_m^{AB})\right)^2.
\end{align}
Taking into account that instead of one, there are $N_m$ bonds ( for example, $N_1=8$ and $N_2=6$ for BCC structure) with atoms in the $m$-th shell, then according to central limit theorem, the local energy in an infinite system follows Gaussian distribution, with the variance given by
\begin{align}
    \sigma(E_\text{loc})^2 = \sum_m \frac{1}{\sqrt{N_m}}\sigma_m^2. \label{variance}
\end{align}
Applying the above formula to the three HEAs, the standard deviation of the local energies are calculated and shown in Tab. \ref{tab:t1}.
\begin{table}[!ht] 
%\footnotesize
\centering
\caption{Standard deviation of local energies for the three HEAs. $\sigma(E_\text{loc})$ is obtained with Eq. \ref{variance}, while $\sigma_\text{fit}(E_\text{loc})$ is from fitting the DFT data.}
\label{tab:t1}
\begin{tabular}{@{}cccc@{}}
\toprule
HEA      & \textup{NbMoTaW}   & \textup{NbMoTaWV}   & \textup{NbMoTaWVi}  \\ \midrule
\textup{$\sigma(E_\text{loc})$ (Ry)} & 0.0026 & 0.0075 & 0.0042  \\
\textup{$\sigma_\text{fit}(E_\text{loc})$ (Ry)}  & 0.0024 & 0.0076 &  \\ \bottomrule
\end{tabular}
\end{table}
This results in Tab. \ref{tab:t1} is in agreement with our previous observation that the chemical interactions in NbMoTaWV is stronger than the more ductile NbMoTaW and NbMoTaWTi. For comparison, we also calculate the standard deviation of averaged energy at different supercell sizes, as shown in Fig. \ref{fig:fluctuations}. The local energy standard deviations can then be obtained from the intercept of the $\log$-$\log$ plot. The results for NbMoTaW and NbMoTaWV, which are the two materials we calculated with up to about 1000-atom supercell, are given in Tab. \ref{tab:t1}. Both results are in excellent agreement with the theoretical values. From the slopes (-0.45 for NbMoTaW and -0.48 for NbMoTaWV) of the lines in Fig. \ref{fig:fluctuations}, we can also see that the standard deviation of the averaged energy indeed decrease approximately as $1/\sqrt{N}$, as required by the central limit theorem.

The local energy fluctuation in HEAs introduces some fundamental challenges to the theory of plasticity. In conventional alloys, the generalized stacking fault energies (SFE) is a very important quantity. For example, the intrinsic SFE determines the separation of partial dislocations in FCC metals, and the profile of the generalized SFEs govern the dislocation core structure, which further affects the motion of dislocations. For systems with small $\sigma(E_\text{loc})$, it is then expected that an averaged SFE should well characterize the system, and can be feed into traditional methods such as Peierls model. If $\sigma(E_\text{loc})$ is large, then new theory should be developed to incorporate the fluctuation effects. Therefore, the magnitude of $\sigma(E_\text{loc})$ can guide us on the validity of traditional theory. On the other hand, in first principles calculations it is often necessary to calculate the averaged energy with very high accuracy \cite{Ding8919, LIU2019107955}, which requires the use of large supercell to reduce the fluctuation (the variance of averaged energy scales as 1/N). Since the EPIs are intrinsic to the elements, we can use these information to estimate the minimum supercell needed for the averaged energy to reach a given precision. 

\section{Acknowledgements}
Xianglin Liu would like to thank Zongrui Pei and Fernando Reboredo for fruitful discussions. The work of X. L. and M. E. was supported by the U.S. Department of Energy, Office of Science, Basic Energy Sciences, Materials Science and Engineering Division. J. Z. was supported by the Laboratory Directed Research and Development Program of Oak Ridge National Laboratory. This research used resources of the Oak Ridge Leadership Computing Facility, which is supported by the Office of Science of the U.S. Department of Energy under Contract No. DE-AC05-00OR22725. 

\bibliographystyle{model1-num-names}
\bibliography{PairInteraction.bib}

%% Authors are advised to submit their bibtex database files. They are
%% requested to list a bibtex style file in the manuscript if they do
%% not want to use model1-num-names.bst.

%% References without bibTeX database:

% \begin{thebibliography}{00}

%% \bibitem must have the following form:
%%   \bibitem{key}...
%%

% \bibitem{}

% \end{thebibliography}

\end{document}